\documentclass[twocolumn,showpacs,showkeys,preprintnumbers]{revtex4}

\usepackage{amsmath,amssymb}

\def\bra{\langle}
\def\ket{\rangle}

\def\qbar{\overline{{\rm q}}}

\def\sbar{\overline{{\rm s}}}

\def\threebar{\overline{{\mathbf{3}}}}

\def\toprule{\hline}
\def\midrule{\hline}
\def\bottomrule{\hline}

\def\Voge{{V_{\rm OGE}}}
\def\Vps{{V_{\rm PS}}}
\def\Vobe{{V_{\rm OBE}}}
\def\Vconf{{V_{\rm conf}}}

\def\aconf{{a_{\rm conf}}}

\def\ohalfp{\smash{${1\over 2}^+$}}
\def\ohalfm{\smash{${1\over 2}^-$}}
\def\thalfm{\smash{${3\over 2}^-$}}

\def \mbf#1{\mbox{\boldmath{$#1$}}}

\def\vecxi{\mbf{\xi}}
\def \veceta{\mbf{\eta}}
\def \vecr{\mbf{r}}
\def \vecR{\mbf{R}}

\def\MeV{{\footnotesize{[MeV]}}}
\def\OGE{{\footnotesize{OGE}}}

\def\lamilamj{(\lambda_i\cdot\lambda_j)}
\def\sigisigj{(\mbf{\sigma}_i\cdot\mbf{\sigma}_j)}

\def\fifj{(\mbf{f}_i\cdot \mbf{f}_j)}

\newcommand{\FRAC}[2]{\leavevmode\kern.1em
  \raise.5ex\hbox{\the\scriptfont0 #1}\kern-.1em
  /\kern-.15em\lower.25ex\hbox{\the\scriptfont0 #2}}

\begin{document}
\setlength{\baselineskip}{15pt}

\title{Pentaquark  as a NK$^*$ bound state with $TJ^P$=0${3\over 2}^-$}
\author{Sachiko Takeuchi}\affiliation{%
Japan College of Social Work, Kiyose, Tokyo 204-8555, Japan}
\author{Kiyotaka Shimizu}\affiliation{%
Department of Physics, Sophia University, Chiyoda-ku, Tokyo 102-8554, Japan
}
\date{\today}


\begin{abstract}
We have investigated negative-parity uudd$\sbar$ pentaquarks 
by employing a quark model 
with the meson exchange and the effective gluon exchange 
as  qq and q$\qbar$ interactions.
The system of five quarks is dynamically solved;
the qq and q$\qbar$ correlations
are taken into account  in the wave function.
The masses of the pentaquarks
are found to be reasonably low.
It is found that the lowest-mass state is $TJ^P$=0\ohalfm\
and the next lowest one is 0\thalfm.
The former is reported to have a large width.
We argue the observed narrow peak corresponds to the latter state.
It is still necessary to introduce an extra attraction to reduce the mass
further by 140 -- 280 MeV to reproduce the observed $\Theta^+$ mass.
Since their level splitting is less than 80 MeV,
the lower level will not become a bound state below the NK threshold
even after such an attraction is introduced.
It is also found that the relative distance of two quarks with the attractive interaction 
is found to be by about 1.2 -- 1.3 times closer than that of the repulsive  one.
The two-body correlation seems important in the pentaquark systems.
\end{abstract}

\maketitle

\section{Introduction}

Since the experimental discovery of the baryon resonance with strangeness
+1, $\Theta$(1540)$^+$ \cite{nakano}, 
many attempts have been performed to describe the peak theoretically\cite{oka}.
To describe this resonance by using a quark model, 
one needs at least five quarks, uudd$\sbar$, 
which is called a pentaquark.
After one year of straggle,
it gradually has become clear that a quark model has difficulties
to explain some of the features of this peak.
Namely, 
(1) the  observed mass is rather low, 
(2) the observed width is very narrow, and
(3) there is only one peak is found, especially no $T$=1 peak  nearby.
In order to reproduce the observed mass, about 100 MeV above the KN threshold,
it is preferred to assign 
the $(0s)^5$ state with the most attractive channel, $TJ^P$=0\ohalfm.
It has been pointed out, however, that 
width of this state would be about 1 GeV  \cite{hosaka}, 
which is far from the observed narrow width, 0.90 MeV \cite{pdg}.
The possibility that the pentaquark with  0\thalfm as well as 0\ohalfp
may be seen as a low-lying peak
was pointed out by several works \cite{oka, inoue, hosaka}.
In this work, we would like to show that 
the pentaquark with 0\thalfm is a promising candidate for the observed peak
by performing a dynamical calculation of the five-quark system with the 
realistic qq and q$\qbar$ interactions.

We employ two kinds of parameter sets for the hamiltonian:
the one is with the one-boson exchange (OBE) as the qq interaction,
the other is with 
the one-gluon exchange (OGE) as well as OBE.
We find that 
the absolute value of the mass is low after a reasonable assumption
for the zero-point energy is introduced, though
it is still necessary to introduce an extra attraction to reproduce the data.
We also find that both of the two parameter sets predict that
the mass of the 0\ohalfm\ state is lower than that of the 0\thalfm\
state.
Their difference, however, is less than 80 MeV.
Thus, the 0\thalfm\ state can be assigned to the observed peak
without forming the 0\ohalfm\ bound state below NK threshold
even after the extra attraction is introduced.

\section{Model}

We have employed a valence quark model. 
 The hamiltonian is taken as:
 \begin{eqnarray}
H_q &=& \sum_i\sqrt{m_i^2+p_i^2}  +v_0 \nonumber \\
&+& \sum_{i<j} \left(\Voge_{ij} + \Vobe_{ij} + \Vconf_{ij}\right).
\end{eqnarray}
The two-body potential term consists of the one-gluon-exchange potential, 
$\Voge$\cite{ruju}, the one-boson-exchange
potential, which consists of the PS and $\sigma$-meson exchange, 
$\Vobe=\Vps+V_\sigma$, and the confinement potential, $\Vconf$, which are defined as:
\begin{widetext}
\begin{eqnarray}
\Voge_{ij} &=&\lamilamj{\alpha_s\over 4}\left\{
\left({1\over r_{ij}}\!-\!{e^{-\Lambda_g r_{ij}}\over r_{ij}}\right)
\!-\! \left({\pi\over 2 m_i^2}\!+\!{\pi\over 2 m_j^2}\!+\!{2\pi\over 3 m_im_j}\sigisigj \!\right)
{\Lambda_g^2\over 4\pi}{e^{-\Lambda_g r_{ij}}\over r_{ij}}
\right\},
\\ 
\Vps_{ij} &=& \frac{1}{3} \frac{g^2}{4 \pi} \frac{m_{\rm m}^2}
{4m_im_j} 
\fifj \sigisigj
\left \{ \frac{e^{-m_{\rm m}r_{ij}}}{r_{ij}}
 -\left( \frac{\Lambda_{\rm m}}{m_{\rm m}} \right)^2
\frac{e^{-\Lambda_{\rm m} r_{ij}}}{r_{ij}} \right \},
\label{OMEP}
\\
V_{\sigma~ij} &=& -\frac{g_8^2}{4 \pi} 
\left \{ \frac{e^{-m_{\rm m}r_{ij}}}{r_{ij}}
\!-\!\frac{e^{-\Lambda_{\rm m} r_{ij}}}{r_{ij}} \right \}~,
\\
{\Vconf}_{ij} &=& 
\begin{cases}
-\lamilamj \; \aconf \;r_{ij} &{\text{\rm ~~(q$\qbar$ and q$^3$ systems)}}\\
\;\displaystyle{4\over 3}\; \aconf \;r_{ij} &{\text{\rm ~~(q$^4\qbar$ systems)}}\\
\end{cases}~~.
\label{eq:conflam}
\end{eqnarray}
\end{widetext}
In $\Voge$, $\alpha_s$ is the strength of OGE, and $\Lambda_g$ is the
form factor introduced  because quarks cannot be considered 
as point-like particles in this picture.
In $\Vps$, $g$ is the quark-meson coupling constant:
 $g=g_8$ for  $\pi$, $K$, and $\eta$ 
and $g=g_0$ for $\eta'$ meson. 
From the asymptotic potential shape, $g_8$ can be obtained from the observed 
 nucleon-pion coupling constant, $g_{\pi {\rm NN}}$\cite{glozm1,shiglo,FS02,FST03}. 
$\mbf{f}$ and 
$\mbf{\sigma}$ are the flavor U(3) generators and 
Pauli spin operators, respectively.
The term proportional to $(\Lambda_{\rm m}/m_{\rm m})^2$ 
is originally the $\delta$-function term;
the form factor for the meson exchange, $\Lambda_{\rm m}$, is also introduced.
$\Lambda_{\rm m}$ is assumed to 
depend on the meson mass $m_{\rm m}$ as:
$\Lambda_{\rm m}=\Lambda_0+ \kappa \;m_{\rm m}$ 
\cite{glozm1,shiglo,FS02,FST03}.

%
%
\begin{table*}[tb]
\caption{Parameter sets.
Each parameter set is denoted by R$\pi$, or Rg$\pi$.
}
\begin{center}
\label{tbl:QMparam}
\def\SPA{\phantom{0}}
\def\SPB{\phantom{$-$0}}
\def\SPC{\phantom{.0}}
\tabcolsep=0.5mm
\begin{tabular}{l@{~~}ccccccccccccccc}
\hline 
Model&Kin& qq int.& $m_{\rm u}$ &$m_{\rm s}$ 
& $\alpha_s$ & $\Lambda_g$ & $\FRAC{$g^2_8$}{$4\pi$}$ 
&$(\!\FRAC{$g_0$}{$g_8$})^2$& $\Lambda_0$ &$\kappa$& 
$m_\sigma$ &$\aconf$ & $V_0$ &$V_0'\!-\!2V_0$\\
ID&&&\MeV&\MeV&&\footnotesize [fm\!\!$^{-1}$]&&&
\footnotesize [fm\!\!$^{-1}$]&&\MeV &\footnotesize [MeV\!/fm] &\MeV &\MeV \\ \hline 
R$\pi^\dag$ & SR & $\pi$ $\sigma$ $\eta$& 
313 & 530 & 0 & - & 0.69\SPA &0& 1.81 &0.92 &675& 170\SPC &$-$378.3 & $-$51.7\\
Rg$\pi$ & SR & \OGE\ $\pi$ $\sigma$ $\eta$ $\eta'$& 
340 & 560 & 0.35 & 3 & 0.69\SPA & 1 & 1.81 & 0.92 & 675 & 172.4 & $-$381.7 &  $-$22.4\\
Graz$^\ddag$ & SR & $\pi$ $\eta$ $\eta'$&
340 & 500 &0 & - & 0.67\SPA &1.34& 2.87 &0.81 &-&172.4 & $-$416\SPC &  $-$39.3\\ \hline 
\end{tabular}
\end{center}
\rule{1.5cm}{0cm}$^\dag$ \cite{FST03}, $^\ddag$\cite{glozm1}  with new $V_0'$. \hfill\rule{0cm}{0cm}
\end{table*}

We have employed two kinds of parameter sets: the set
with $\Vobe$ but not $\Voge$ (chiral model, R$\pi$ in the following) and 
the set with $\Voge$ and $\Vobe$ (Rg$\pi$) as shown in Table \ref{tbl:QMparam}.
 For reference, we also employ the parameter set given by
 the Graz group \cite{glozm1}.

As for the confinement potential for 
pentaquarks, we replace the factor $\lamilamj$ 
by its average value as shown in eq.\ (\ref{eq:conflam}).
This modified potential 
gives the same value as that given by the original  confinement for the orbital $(0s)^5$ state.
This replacement enables us to remove all the scattering states
and to investigate only tightly bound states, which will appear
as narrow peaks.
It is based on the idea of the flux tube model;
there the configurations 
where gluonic flux tubes bind all the five quarks can be distinguished
from those of a baryon with a meson.
After the coupling of the scattering states with an original confinement 
potential, some of the states we find will melt away into the continuum \cite{hiyama}.
Later we discuss which states disappear by comparing the masses of 
the pentaquarks and the baryon-meson states.

We take the zero-point energy, $v_0$, as:
\begin{eqnarray}
{v_0} &=& 
\begin{cases}
\;V_0' & \text{~~q$\qbar$ systems}\\
\;3 V_0 & \text{~~q$^3$ systems}\\
\;6 V_0 & \text{~~q$^4\qbar$ systems}
\end{cases}~~.
\label{eq:Vzero}
\end{eqnarray}
The zero-point energy of the pentaquark 
is taken to be twice as large as that of the q$^3$ systems.
It is motivated by the result of lattice QCD calculation, which indicates
that the qq-$\sbar$-qq  type gluon configuration  
is favored for a pentaquark \cite{Suganuma,enyo};
namely,  two Y-shapes which are 
connected by the $\sbar$ quark 
 gives the lowest energy.
The value of the zero-point energy itself, however, is not uniquely
determined in this kind of empirical models.
Our main concern here is the level splitting of the states,
though we believe the above assumption is not very far from reality. 
\bigskip

The wave function we employ is written as:
\begin{eqnarray}
\lefteqn{\psi_{TSL}(\vecxi_A,\vecxi_B,\veceta,\vecR) = 
\sum_{i,j,n,m,\alpha,\alpha',\lambda} c_{ijnm}^{\alpha \alpha' \lambda}\;
\mathcal{A}_{q^4}\;
}\nonumber \\
&\times&
\phi_{\rm q^2}(\alpha,\vecxi_A;u_i)\;\phi_{\rm q^2}(\alpha',\vecxi_B;u_j)\;
\psi(\lambda,\veceta;v_n)\Big|_{TSL}\;\nonumber \\
&\times&\chi_{\sbar}(\vecR;w_m)
\end{eqnarray}
where $\mathcal{A}_{q^4}$
is the antisymmetrization operator over the four ud-quarks, and
$\vecxi_A$, $\vecxi_B$, $\veceta$ and $\vecR$
are the coordinates defined as:
\begin{eqnarray}
\vecxi_A &=& \vecr_1-\vecr_2 ~~{\rm and}~~
\vecxi_B = \vecr_3-\vecr_4 \\
\veceta &=& (\vecr_1+\vecr_2 -\vecr_3-\vecr_4 )/2\\
\vecR &=& (\vecr_1+\vecr_2 +\vecr_3+\vecr_4)/4-\vecr_{\sbar}
\end{eqnarray}
$\phi_{\rm q^2}(\alpha,\vecxi;u)$ is the wave function for a qq pair 
with 
the size parameter $u$:
\begin{eqnarray}
\phi_{\rm q^2}(\alpha,\vecxi;u)&=& 
\varphi_\alpha
\;\exp\left[-\dfrac{\xi^2}{4 u^2}\right] 
\end{eqnarray}
where the quantum number $\alpha$ stands for one of the four relative $S$-wave quark pairs:
$(TS)C$ = (00)$\threebar$, (01)${\mathbf{6}}$, (10)${\mathbf{6}}$, and (11)$\threebar$.
The relative wave function between two quark pairs, $\psi(\lambda,\veceta;v)$, and 
the wave function between the four-quark cluster and the $\sbar$ quark, $\chi_{\sbar}(\vecR;w)$, are
taken as:
\begin{eqnarray}
\psi(\lambda,\veceta;v)&=&
\exp\left[-\dfrac{\eta^2}{2 v^2}\right] 
\\
\chi_{\sbar}(\vecR;w)
&=&\exp\left[-\dfrac{2 R^2}{5 w^2}\right].
\end{eqnarray}

The gaussian expansions are taken as geometrical series:
$u_{i+1}/u_i$=$v_{n+1}/v_n$=2 and $w_{m+1}/w_m$=1.87.
We take 6 points for $u$ (0.035 -- 1.12fm),
4 points for $v$ (0.1 -- 0.8fm), and 3 points for $w$ (0.2 -- 0.7fm).
Since we use a variational method, the obtained masses are the upper-limit.
They, however, converge rapidly; the mass may reduce more, but probably only by several MeV.

\section{mass spectrum}

The masses of q$\qbar$, q$^3$,  and q$^4\qbar$ systems  are shown in Table \ref{tbl:mass}.
N, $\Sigma$, and $\Delta$ masses of R$\pi$ and Graz parameter sets  
were given in refs.\ \cite{glozm1, FST03}.

It is very difficult for a constituent quark model to describe the Goldstone bosons:
they need a collective mode, which is constructed by the superposition of (q$\qbar$)$^n$. 
Also, it is hard to justify the models with the kaon-exchange interaction between
quarks to describe a kaon.
We do not push the model to give the correct kaon mass.
After fitting $\rho$-meson mass by adjusting $V_0'$ in the eq.\ (\ref{eq:Vzero}), we 
 use K$^*$ mass as a reference of the threshold.

Contrary to the q$\qbar$ systems, 
we have more satisfactory results for the q$^3$ baryons. The masses of
the $S$-wave ground states are well reproduced.
Each parameter set 
was taken so as to approximately reproduce N, $\Delta$, and $\Sigma$ masses.
Though we do not recite other baryon masses, the octet baryon masses are
predicted within less than 25 MeV error in the Graz parameter set, 
41 MeV in R$\pi$, and 13 MeV in Rg$\pi$ parameter set.
The decuplet baryon masses are predicted 
within less than 14 MeV in the Graz parameter set, 
93 MeV in R$\pi$, and 5 MeV in Rg$\pi$ parameter set.
The R$\pi$ parameter set tends to overestimate
the strange baryons. The level splittings themselves are not very far from the observed
values \cite{FST03}.

%
%
\begin{table}[btp]
\caption{Masses of mesons, baryons, N+K$^*$ threshold, 
and pentaquarks of the $TJ^P$ state for each parameter set.
All masses are given in MeV.}
\label{tbl:mass}
\begin{center}
\def\SPC{\phantom{.0}}
\begin{tabular}{l@{~~}ccccccc@{~~}ccc}
\toprule
& N & $\Sigma$ & $\Delta$ & K & $\rho$ & K$^*$ &  NK$^*$ & \multicolumn{3}{c}{Pentaquarks}\\
&&&&&&&& 0\ohalfm & 0\thalfm & 1\ohalfm \\[1.2ex] \midrule
R$\pi$     & 941 & 1191 &1261 & 900 &776 & 928 &1869 & 1730 & 1762 & 1765\\
Rg$\pi$  &938 & 1192 &1231 & 814 & 776 & 908 & 1846 &  1603 & 1682 & 1697\\
Graz  & 937 &1178& 1239 & 890 & 776 & 888 &1825 &  1815&1824 & 1835\\ \midrule
Exp.$^\S$  & 939 & 1193 &1232 & 494 & 776 &  892 &1831& 
\multicolumn{3}{c}{1540}\\ \bottomrule
\end{tabular}
\end{center}
$^\S$Ref.\ \cite{pdg} \hfill\rule{0cm}{0cm}
\end{table}

We have solved 
the system of the pentaquarks using the method described in the previous section.
The masses of the pentaquark with $TJ^P$= 0\ohalfm, 0\thalfm, and 1\ohalfm are shown
in the Table  \ref{tbl:mass}.

As for the chiral quark models,
it is known from the group theoretical consideration 
that
the  ($TS$)=(01) and (10) states are the lowest two among the q$^4$ $S$-wave systems 
with the flavor-spin interaction \cite{carlson, stancu, ST04}.
Since there is no pion-exchange between u or d  and $\sbar$ quark,
 the  three states,
 $(TS)J^P$= (01)\ohalfm, (01)\thalfm, and (10)\ohalfm, are essentially degenerated.
 In our case, R$\pi$ and Graz  parameter sets are the chiral models.
The mass difference of these three levels is 20--34 MeV in these parameter sets.

 Both of the $\Voge$ and $\Vobe$ are included in the Rg$\pi$ parameter set.
Because of $\Voge$ has non-vanishing spin-spin interaction 
between the q$^4$ cluster and $\sbar$-quark,  
the splitting between (01)\ohalfm\ and  \thalfm\ is
much larger in Rg$\pi$ than that of the chiral model:
it is 71 MeV for the Rg$\pi$ parameter set 
whereas it is 32 MeV for $R\pi$,  or 9 MeV for the Graz parameter set.

The absolute values of the pentaquark mass are from 1603 to 1835 MeV.
Each of the states is below the NK$^*$ threshold except for one
exception, 1\ohalfm\ of Graz parameter set.
Since the assumption we made for the zero-point energy
has large ambiguity, we do not conclude that they {\it are} the 
pentaquark mass.  
More attraction is necessary to reproduce the observed pentaquark mass.
\bigskip

Let us discuss 
which of the above levels should be observed as a peak.
It is known that
for the ($TS$)=(01) and (10) state, there is only one spin-flavor-color configuration which 
can be combined to the orbital [4] symmetry \cite{carlson, stancu, ST04}.
This means that a pentaquark which includes the above q$^4$ states 
can couple to  the relative $S$-wave meson-baryon systems strongly.

Suppose that
a peak is observed only when the level is below the `$S$-wave threshold',
by which we mean the mass of the meson-baryon system 
which can form the concerning $TJ^P$ state
with relative $S$-wave.
For example, the $S$-wave threshold of
the $TJ^P$=0\ohalfm\ and 1\ohalfm\ states 
is $m_{\rm N}+m_{\rm K}$ while that of 
0\thalfm\ is $m_{\rm N}+m_{{\rm K}^*}$. 
Then, the levels of 0\ohalfm\ and 1\ohalfm\
disappear if they are higher than the
NK threshold while the level of 0\thalfm\ may be seen
if it is lower than the NK$^*$ threshold.
Also,
1\thalfm\ and 2\thalfm\ disappear if they are higher than the $\Delta$K threshold.

As seen in Table \ref{tbl:mass}, 
the $TJ^P$=0\ohalfm\ and 1\ohalfm\ states  are above the NK threshold in our present work.
Thus these two are probably not observed.
On the other hand, the mass of the 0\thalfm\ state
is below the  NK$^*$ threshold.
Because this level has to decay to the relative $D$-wave NK system 
by the tensor term of the interaction,
this level may be seen as a peak.
To investigate the situation quantitatively,
one needs to perform, {\it e.g.}\
a resonating-group-method calculation for q$^4\qbar$ systems \cite{TKN},
by which the width of the state can be obtained.
This we will investigate elsewhere.

To assign the 0\thalfm\ state to the observed peak,
it is still necessary to introduce extra attraction by 140 -- 280 MeV.
It is reported that there are  other sources which contribute the 
absolute mass. For example, the instanton induced interaction, 
which should be taken into account
to reproduce the $\eta$-$\eta'$ mass difference, gives a universal
two-body attraction and a three-body repulsion\cite{shinozaki}.

The level splitting 
between the lowest two states, $TJ^P$=0\ohalfm and 
0\thalfm,   is less than 80 MeV.
So, the lowest state will not become a bound state as the 
extra attraction is introduced so that the
0\thalfm\ state becomes 100 MeV above the NK threshold.
Other states which can be combined to the orbital [4] symmetry
are known to have a higher mass from the  discussion based on the
group theory \cite{ST04}.
It is also pointed out that one of the positive-parity pentaquarks, 
 0\ohalfp\ state,
may be assigned to the observed single peak.
Actually, this level can be as low as the negative-parity state  \cite{ST04}.
It is argued, however, that the width of this state seems still wider than the
observed one \cite{hosaka}.
 This 0\thalfm\  pentaquark 
seems more appropriate candidate of the observed single peak.

\section{Roles of the qq correlation}

%
%
\begin{table}[bt]
\caption{The number of quark pairs with the quantum number $T_2S_2$, $N_{T_2S_2}$,
and the size of the pairs, $r_{T_2S_2}$, in fm.
For definition, see text.}
\label{tbl:rr}
\begin{center}
\def\SPA{\phantom{0}}
\def\SPB{\phantom{$-$0}}
\def\SPC{\phantom{.0}}
\begin{tabular}{cccrrrrrrrr}
\toprule
&&&\multicolumn{4}{c}{$(T_2S_2)$ qq pair}
\\
&($TS$)$J^P$ & &(00) & (01) & (10) & (11)  
\\ \midrule
R$\pi$
& (01)\ohalfm & 
$N_{T_2S_2}$ & 1.49&  1.51&  0.51&  2.49
\\
&&$r_{T_2S_2}$ & 0.53 & 0.70 & 0.68 & 0.62 
\\[1.1ex]
& (01)\thalfm & 
$N_{T_2S_2}$ & 1.49&  1.51&  0.51&  2.49
\\
&&$r_{T_2S_2}$  & 0.53 & 0.72 & 0.69 & 0.63 
\\[1.1ex]
& N &
$N_{T_2S_2}$ & 1.48 & 0.02 & 0.02 & 1.48
\\
&&$r_{T_2S_2}$ & 0.50 & 0.65 & 0.65 & 0.56 
\\ \midrule
Rg$\pi$
& (01)\ohalfm & 
$N_{T_2S_2}$ & 1.49&  1.51&  0.51&  2.49
\\
&&$r_{T_2S_2}$  & 0.56 & 0.69 & 0.68 & 0.64 
\\[1.1ex]
& (01)\thalfm & 
$N_{T_2S_2}$ & 1.49&  1.51&  0.51&  2.49
\\
&&$r_{T_2S_2}$ & 0.57 & 0.73 & 0.71 & 0.66 
\\[1.1ex]
&N &
$N_{T_2S_2}$ & 1.49	&0.01	&0.01	&1.49 
\\
&&$r_{T_2S_2}$  & 0.55 & 0.76 &0.76 & 0.62
\\ \midrule
SU(3)
& (01)\ohalfm & 
$N_{T_2S_2}$ &1.5 & 1.5 & 0.5 & 2.5
\\
& (01)\thalfm & 
$N_{T_2S_2}$ &1.5 & 1.5 & 0.5 & 2.5
\\
&N &
$N_{T_2S_2}$ &1.5 & 0 & 0 & 1.5 \\
\bottomrule
\end{tabular}
\end{center}
\end{table}

Except for the confinement force, all the interaction terms
are short-ranged in the quark model.
Thus, when the deformation by the quark-quark correlation is introduced in the model,
quark pairs where the interaction 
is attractive become more tightly bound  while
those with repulsion tend to stay apart from each other.
Then an attractive pair may behave like a single particle;
this is  the qq correlation which motivates the diquark models \cite{JW,KL}.
We have looked into how much the qq correlation is developed 
in our full calculation by checking the size of each quark pairs.

In Table \ref{tbl:rr}, we show the number of quark pairs 
with specific quantum numbers and the size of that pairs.
The number of the pairs with the quantum number $T_2S_2$, $N_{T_2S_2}$,
and the size of the pairs, $r_{T_2S_2}$, are
defined  by using the projection operator $P_{ij}^{(T_2S_2)}$ as:
\begin{eqnarray}
N_{T_2S_2} &=& \left\bra \sum_{i>j} P_{ij}^{(T_2S_2)} \right\ket \\
r_{T_2S_2} &=& 
\sqrt{\left\bra \sum_{i>j} P_{ij}^{(T_2S_2)}r_{ij}^2 \right\ket \Big/N_{T_2S_2}}~~.
\end{eqnarray}

The number of quark pairs, $N_{T_2S_2}$, obtained by the full calculation
is not very different from that of the group classification, 
as was also found in the nucleon case\cite{FS02}. 
The contribution from each pair, however, can be different.
The size of quark pairs is large when the interaction is repulsive 
while it becomes small for the attractive
pairs. The ratio is about 1.2 -- 1.3.
We also find that the qq correlation in the pentaquarks 
have similar size to that in the nucleon.

\section{Summary}

We have investigated the negative-parity uudd$\sbar$ pentaquarks 
by employing a quark model.
The system for the five quarks is dynamically solved;
the effects of  qq or q$\qbar$ correlations on the wave function 
are taken into account.
The model has realistic  qq and q$\qbar$ interactions:
the meson exchange for the chiral models, and both of the meson 
and  the effective gluon exchange for the other parameter set, Rg$\pi$. 

It is found that the masses of the pentaquarks
are reasonably low, though it is still necessary to introduce an extra attraction 
to reduce the mass
further by 140 -- 280 MeV to reproduce the observed $\Theta^+$ mass.
The pentaquark of the lowest mass is found to be $TJ^P$=0\ohalfm.
The next lowest is 0\thalfm; 
we argue the observed peak corresponds to the latter state
because the width can be narrow for this state.
Since the level splitting of these two states is no more than 80 MeV,
the lower level will not become a bound state below the NK threshold
even if we introduce the extra attraction
so that  the mass of the upper state become as low as the observed one.
The lower level will melt into the continuum after the coupling 
to the meson-baryon states is introduced.  
 
It is also found that the size of quark pairs with the attractive interaction 
is found to be by about 1.2 -- 1.3 times closer than that of the repulsive  one.
The two-body correlation seems important in the pentaquark systems.

\begin{acknowledgements}
This work is supported in part by a Grant-in-Aid for Scientific Research
from JSPS (No.\ 15540289).
\end{acknowledgements}

\end{document}